\documentclass[11pt,halfline,a4paper]{article}

\usepackage[utf8]{inputenc}
\usepackage[T1]{fontenc}
\usepackage{amsmath}
\usepackage{amsfonts}
\usepackage{amssymb}
\usepackage{mathrsfs}
\usepackage{glossaries}

\usepackage{fancyhdr}
\usepackage{blindtext}

\usepackage{MnSymbol}

\usepackage[cal=boondox,scr=boondoxo]{mathalfa}

\usepackage{float}
\usepackage{fancybox,graphicx}
\usepackage{subfigure}
\usepackage{subcaption}
\usepackage{caption}
\usepackage{color}
\usepackage{authblk}
\usepackage{listings}
\usepackage[colorlinks=true, urlcolor=blue]{hyperref}

\usepackage{hyperref}

\newcommand{\doi}[1]{{doi:~\href{https://doi.org/#1}{\nolinkurl{#1}}}\rmFullStop}

\newcommand*{\rmFullStop}{\rmifnextchar{.}{}{}}

\makeatletter
\newcommand{\rmifnextchar}[3]{%
  \begingroup
  \ltx@LocToksA{\endgroup#2}%
  \ltx@LocToksB{\endgroup#3}%
  \ltx@ifnextchar{#1}{%
    \def\next{\the\ltx@LocToksA}%
    \afterassignment\next
    \let\scratch= %
  }{%
    \the\ltx@LocToksB
  }%
}
\makeatother 

\usepackage{accents}
\usepackage[titletoc,title]{appendix}
\usepackage{cite}
\usepackage[dvipsnames]{xcolor}
\usepackage[numbers,sort&compress,super]{natbib}

\usepackage{mathtools}

\usepackage[top=2in, bottom=1.5in, left=1in, right=1in]{geometry}

\usepackage{stackengine}

\newcommand{\eg}[1]{\textit{e.g.,}}
\newcommand{\ie}[1]{\textit{i.e.,}}
\newcommand{\icode}[1]{\texttt{#1}}

\makeglossaries

\newacronym{API}{API}{application programming interface}
\newacronym{DSL}{DSL}{domain-specific language}
\newacronym{HTTP}{HTTP}{Hypertext Transfer Protocol}
\newacronym{IDL}{IDL}{interface definition language}
\newacronym{JSON}{JSON}{JavaScript object notation}
\newacronym{URI}{URI}{Uniform Resource Identifier}
\newacronym{UML}{UML}{Unified Modeling Language}
\newacronym{HL7}{HL7}{Health Level Seven}
\newacronym{FHIR}{FHIR}{Fast Healthcare Interoperability Resources}
\newacronym{PAWS}{PAWS}{Pediatric Apple Watch Study}
\newacronym{ECG}{ECG}{electrocardiogram}

\glsunset{HTTP}
\glsunset{JSON}
\glsunset{URI}

\glsdisablehyper

\title{Spezi Data Pipeline: Streamlining FHIR-based Interoperable Digital Health Data Workflows
}

\author[1,2,3]{Vasiliki Bikia, \href{https://www.orcid.org/0000-0002-4660-1560}{0000-0002-4660-1560}}
\author[3]{Paul Schmiedmayer, \href{https://www.orcid.org/0000-0002-8607-9148}{0000-0002-8607-9148}}
\author[4]{Aydin Zahedivash, \href{https://orcid.org/0000-0001-6835-1139}{0000-0001-6835-1139}}
\author[3]{Lauren Aalami, \href{https://orcid.org/0009-0007-7132-5362}{0009-0007-7132-5362}}
\author[3]{Adrit Rao, \href{https://orcid.org/0000-0002-0780-033X}{0000-0002-0780-033X}}
\author[3]{Vishnu Ravi, \href{https://www.orcid.org/0000-0003-0359-1275}{0000-0003-0359-1275}}
\author[3]{Matthew Turk, \href{https://www.orcid.org/0000-0000-0000-0000}{0000-0000-0000-0000}}
\author[4]{Scott R. Ceresnak, \href{https://orcid.org/0000-0002-9473-0105}{0000-0002-9473-0105}}
\author[3]{Oliver Aalami \href{https://www.orcid.org/0000-0002-7799-2429}{0000-0002-7799-2429}}

\affil[1]{Stanford University, Department of Biomedical Data Science, Stanford, CA, USA}
\affil[2]{Stanford University, Stanford Institute for Human-Centered Artificial Intelligence, Stanford, CA, USA}
\affil[3]{Stanford University, Stanford Mussallem Center for Biodesign, Stanford, CA, USA}
\affil[4]{Stanford University, Lucile Packard Children's Hospital, Department of Pediatrics, Pediatric Cardiology, Stanford, CA, USA}

\begin{document}

\maketitle

\begin{abstract}
The increasing adoption of digital health technologies has amplified the need for robust, interoperable solutions to manage complex healthcare data.  
We present the Spezi Data Pipeline, an open-source Python toolkit designed to streamline the analysis of digital health data, from secure access and retrieval to processing, visualization, and export. The Pipeline is integrated into the larger Stanford Spezi open-source ecosystem for developing research and translational digital health software systems.  
Leveraging \gls{HL7}~\gls{FHIR}-based data representations, the pipeline enables standardized handling of diverse data types--including sensor-derived observations, \gls{ECG} recordings, and clinical questionnaires--across research and clinical environments.
We detail the modular system architecture and demonstrate its application using real-world data from the \gls{PAWS} at Stanford University, in which the pipeline facilitated efficient extraction, transformation, and clinician-driven review of Apple Watch \gls{ECG} data, supporting annotation and comparative analysis alongside traditional monitors.  
By reducing the need for bespoke development and enhancing workflow efficiency, the Spezi Data Pipeline advances the scalability and interoperability of digital health research, ultimately supporting improved care delivery and patient outcomes.
\end{abstract}

\glsresetall

\section{Introduction}

In a rapidly evolving healthcare landscape, the progressive digitization of data presents a transformative opportunity to enhance clinical outcomes and operational efficiencies.
Amidst rising healthcare expenditures and suboptimal outcomes in the United States, there is a pressing need for innovations that align with the "triple aim" of healthcare improvement: enhancing the individual experience of care, improving the health of populations, and reducing per capita cost \cite{berwick2008triple}.
Digital health technologies--including mobile applications, wearable biometric sensors, telehealth platforms, and computational analytics--represent promising modalities for achieving these objectives through optimized care delivery processes and augmented clinical decision support \cite{witt2019windows, awad2021connected, bert2014smartphones, goel2023mobile}. 

Realizing the full potential of digital health, however, requires overcoming significant challenges in data management. Integrating digital health data from mobile applications and connected devices into research workflows poses significant challenges due to fragmented storage systems, proprietary data formats, and inconsistent access protocols. Retrieving and harmonizing these data sources often requires substantial technical effort and domain expertise, particularly to ensure data integrity and analytical consistency. For data science and biomedical research applications, maintaining high data quality is critical, as inconsistencies or formatting issues can distort results and impair reproducibility.

Standardizing health data remains a critical challenge, particularly where operational systems and research workflows intersect. Health data are inherently heterogeneous, encompassing device-generated metrics, patient-reported outcomes, and clinical records-each with distinct formats, resolutions, and metadata requirements.

Standards such as Open mHealth \cite{OpenmHealth_Schemas}, HL7 \gls{FHIR} \cite{fhir_spec}, and the OMOP Common Data Model (CDM) \cite{OHDSI_CommonDataModel} have enabled more consistent representations of health data and facilitated seamless data exchange in clinical settings. However, their integration into research infrastructure is yet to match the urgency of the need, limiting their impact on scalable and reproducible biomedical studies. Bridging this gap is essential: robust, data-driven research increasingly depends on access to high-quality, harmonized data aligned with evolving interoperability standards. Without such alignment, researchers are forced to rely on ad hoc pipelines and manual harmonization efforts, which are error-prone and difficult to scale. Broader adoption of these standards in research environments offers a unique opportunity to standardize data handling practices, simplify integration and interpretation, and accelerate advances in biomedical science. 

However, a critical layer of infrastructure is still missing--one that would allow researchers to fully leverage these standardized formats. Without tools for structured access, transformation, and visualization, the potential of interoperability remains underrealized.

Preparing raw health data for analysis involves multiple complex steps, including filtering, temporal alignment, feature extraction, and normalization. The challenge is even more pronounced when working with high-dimensional data such as electrocardiogram (ECG) recordings or longitudinal survey responses, which require both technical fluency and domain expertise to analyze meaningfully. Although major cloud providers offer robust APIs-such as Google Cloud Healthcare API~\cite{cloud_healthcare_api} and Azure Health Data Services~\cite{azure_health_data_services} - these solutions are often tightly coupled to specific platforms, require significant customization, and may lack the modularity needed for integration into diverse research workflows. Additionally, while platforms such as Cortex~\cite{cortex2024} have demonstrated successful reuse of digital health data across studies, many lack integration with clinical data standards like \gls{FHIR}, limiting their ability to support scalable, interoperable research aligned with broader healthcare infrastructure.

The Stanford Spezi ecosystem addresses these limitations, providing an open-source, standards\-based foundation for digital and mobile health research \cite{schmiedmayer2025spezi}. The ecosystem's data collection components are cloud-agnostic and support seamless integration of data stored in standardized formats, with strong emphasis on HL7 \gls{FHIR}-based representations. To close the gap between standardized data storage and usable research outputs, we introduce a new component of the ecosystem: the Spezi Data Pipeline--a lightweight, extensible toolkit for extracting, processing, and analyzing digital health data in a reproducible and scalable manner. This Python-based framework is designed to streamline the entire digital health data lifecycle-encompassing access, retrieval, restructuring, analysis, exploration, and export-while adhering to \gls{FHIR} standards for enhanced interoperability. By providing structured access to hierarchical health data and transforming
it into flat, analysis-ready formats, the Pipeline lowers the barrier to working with complex, structured health data and enhances the ability to generate repeatable insights across studies.

To evaluate the Pipeline's performance, we present results from its integration into the real-world \gls{PAWS} at Stanford University. We demonstrate the Pipeline's ability to handle large volumes of mobile health data and efficiently transform them into structured formats ready for interpretation and analysis. 
Ultimately, by providing a standardized, modular solution, the Spezi Data Pipeline aims to reduce the need for bespoke development, improve efficiency, and enhance the reliability of health data management--facilitating improved patient outcomes, reducing operational costs, and enabling scalable, high-quality healthcare delivery.
\section{Methods}

The Spezi Data Pipeline is an MIT-licensed~\cite{spezi_data_pipeline} Python toolkit for standards-based digital health data management within the open-source Stanford Spezi ecosystem~\cite{schmiedmayer2025spezi}.
The Pipeline integrates existing SDKs and libraries into a customizable, modular, FHIR protocol-adherent package with the goal of enabling consistent, secure data handling and interoperability.
In this paper, we detail the Spezi Data Pipeline's design, including its utility in calculating patient activity indices and deriving risk scores from common clinical questionnaires.
We then evaluate the Pipeline's in situ applicability, discussing its implementation in the ongoing Pediatric Apple Watch Study at Stanford University.

The Spezi Data Pipeline package is available to install as a dependency in integrated development environments (IDEs) (\eg, Jupyter Notebook, Google Colaboratory), as well as in standalone Python scripts or larger data processing pipelines. This facilitates seamless integration into a wide range of workflows and enhances the efficiency and reliability of digital health data management.
The code, documentation, and usage examples are hosted on GitHub \cite{spezi_data_pipeline}, enabling easy access and collaboration.

\subsection{Spezi Data Pipeline}

\subsubsection{Supported Resource Types}

\begin{figure}[h!]
  \centering
  \includegraphics[width=\linewidth]{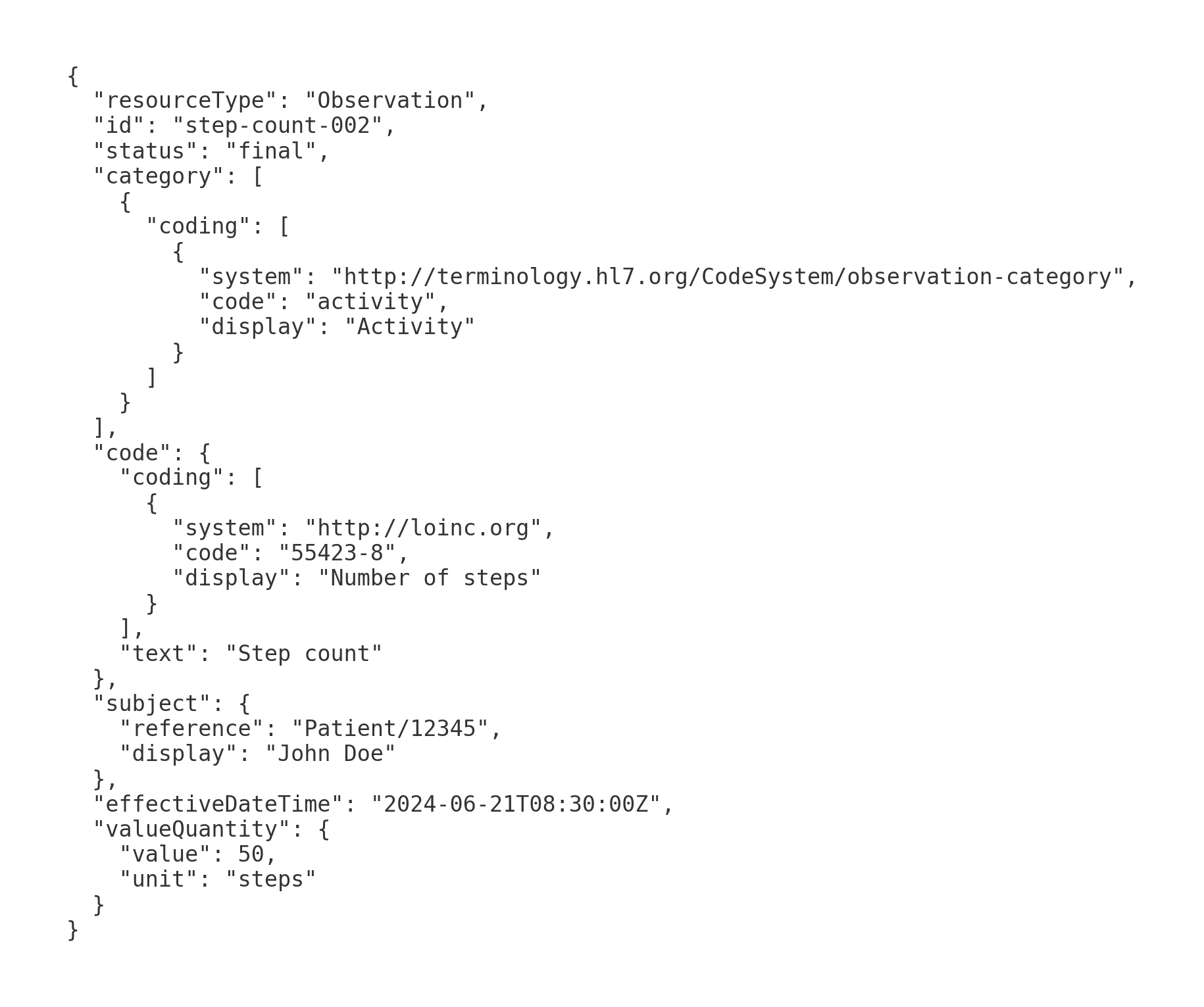}
  \caption{FHIR observation for step count data.}
  \label{fig:fhir_step_count}
\end{figure}

The Spezi Data Pipeline supports many resource data types collected via smartphones and their connected devices, including heart rate, blood pressure, heart rate variability (HRV), electrocardiogram recordings (ECGs), clinical questionnaire responses, and metrics pertaining to activity and nutrition.
Such data are henceforth collectively referred to as "personal health sensor observations," or, simply, "observations."
The Spezi Data Pipeline is designed to process a wide range of observations, enabling their collection and analysis across diverse digital health projects and applications.
Observations are managed using Logical Observation Identifiers Names and Codes (LOINC) codes or other coding standards (such as Apple Health identifiers), providing a standardized method for data handling and classification.
A sample FHIR-compliant observation instance is presented in Figure \ref{fig:fhir_step_count}.

The Spezi Data Pipeline supports the manipulation and analysis of \icode{StructuredDataCapture} specification-adherent clinical questionnaire responses \footnote{https://build.fhir.org/ig/HL7/sdc/}.
Supported surveys include the Social Support Questionnaire (SSQ) \cite{sarason1983assessing}, the Patient Health Questionnaire (PHQ-9) \cite{kroenke2001phq}, and the Walking Impairment Questionnaire (WIQ) \cite{jg1990evaluation}.
Through efficient handling and analysis of clinical questionnaire responses, the Pipeline enables the rapid extraction of actionable insights from patient-reported data, facilitating responsive patient assessment and personalized treatment planning.

\subsubsection{System Design}

\begin{figure}[h!]
  \centering
  \includegraphics[width=0.93\linewidth]{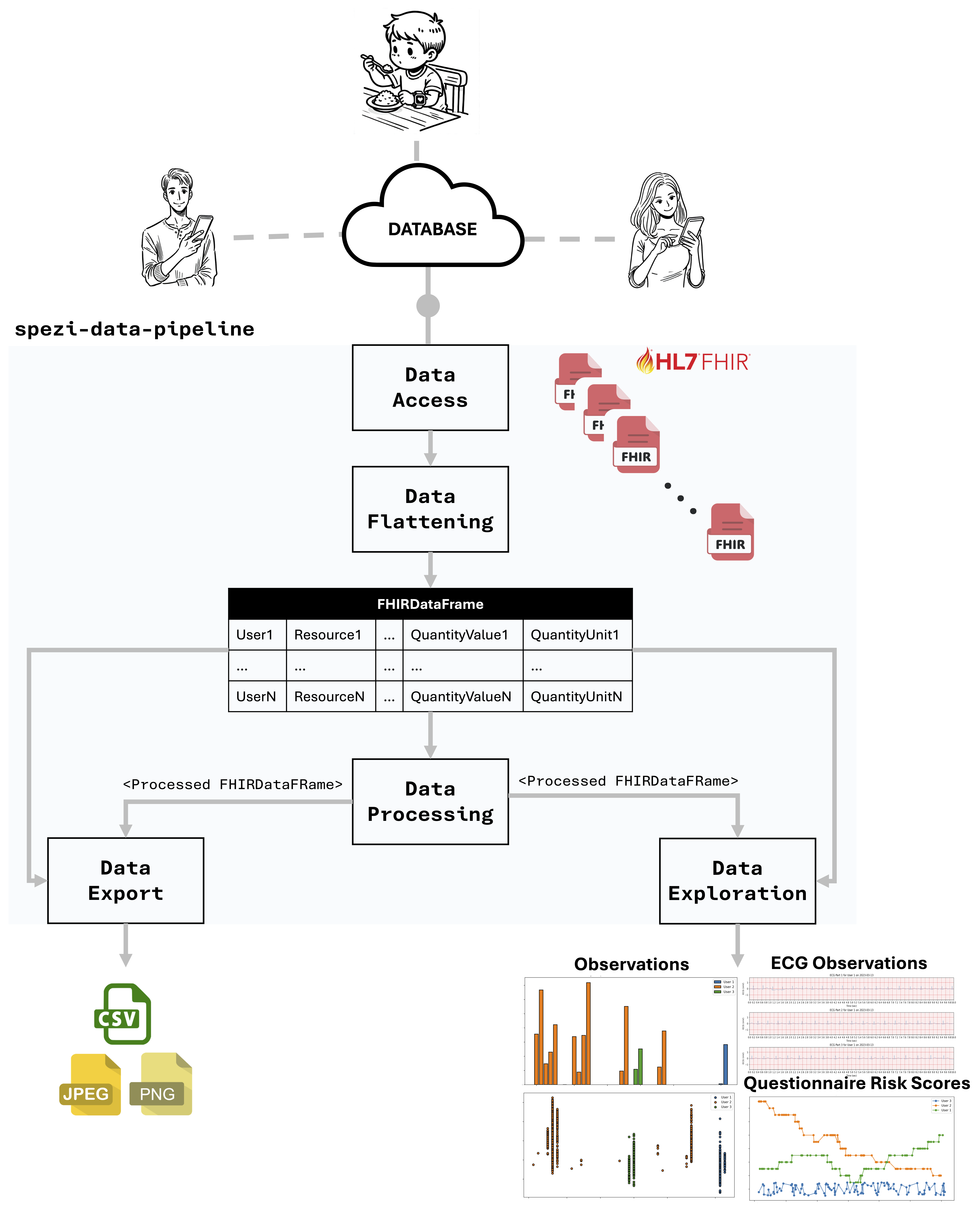}
  \caption{System Design: This diagram depicts the process of obtaining data access from the server, passing through the system modules, and generating outputs. The software is designed to operate within a Python notebook environment.}
  \label{fig:system_design}
\end{figure}

The Spezi Data Pipeline consists of five core modules: Data Access, Data Flattening, Data Processing, Data Exploration, and Data Export (see Figure \ref{fig:system_design}). These modules interact with each other, serving collectively as a comprehensive toolbox for end-to-end data extraction. Importantly, modules can also function as separate components for specific operations in a given system, ensuring modularity and flexibility in the system's design.

\textbf{Data Access.} The Data Access module establishes a connection to the server, functioning as the core component for retrieving information stored in the cloud. This implementation can leverage existing Python SDKs (e.g. Firebase \cite{firebase2024}) for accessing data stored on Google Firebase Firestore, and provides flexibility for developers to set up their own connections to custom cloud platforms or data storage mechanisms.

The module requires data to be stored in the FHIR format to ensure interoperability, and relies on the \icode{fhir.resources} Python package \cite{fhir_resources_package} with built-in data validation for tools and classes for all FHIR resources defined in the FHIR specification \cite{fhir_spec}.
The package enables efficient creation and manipulation of FHIR resources in Python, serving as a well-defined "contract" to ensure data is properly saved and structured.

Once FHIR specification compliance is assured, the Data Access module permits a complete or partial download of the dataset.
Queries for different data types are implemented using standardised coding systems (\eg, LOINC).
After the selected FHIR resources are accessed and retrieved, the raw, nested FHIR resources stored in the cloud are transformed into FHIR resource objects using the \icode{fhir.resources} package.

\textbf{Data Flattening.} Transforming stored data into FHIR resource objects enhances data integrity, but the resulting hierarchical format can complicate data exploration and analysis.
The Data Flattening module is a vital part of the Spezi Data Pipeline, providing tools and classes to convert FHIR data from its native structure into a flattened, tabular \icode{Pandas.DataFrame} format for simplified analysis and modeling.
This tool implements an enhanced \icode{Pandas.DataFrame} called \icode{FHIRDataFrame}, which retains all resource-related metadata.

Through use of this module, data stored in the FHIR format is decomposed and restructured such that each of the resulting data tables contains entries that correspond to a specific FHIR resource.
Each table's columns differ based on the data type (\eg, observation, questionnaire response). For observations, each table includes entries for the user's identifier, resource identifier, quantity name, unit, value, LOINC code, display name, device-associated code (if any), and time and date of the entry.
For \gls{ECG} observations, additional columns include number of measurements, sampling frequency, \gls{ECG} classification, heart rate (with unit), and the \gls{ECG} recording details.
For questionnaire responses, each output table contains columns for the user's identifier, resource identifier, authored date, questionnaire title, question identifier, question text, answer code, and answer text.
Codified questionnaire prompts and responses (in the form of identifiers and codes) in the flattened table should be be mapped to their actual text representations to facilitate meaningful analysis and interpretation.

The Data Flattening tool can additionally be used to extract a comprehensive list of users and their relevant identifying and demographic information, enabling the mapping of user identifiers to their data for subsequent analysis at a later stage.

\textbf{Data Processing.} The Data Processing module contains tools for filtering, selecting, and processing data, utilizing custom classes and methods to filter outliers, aggregating daily data (\eg, summing or averaging), and selecting data within particular windows of time or for specific users.
These functionalities offer flexibility in tailoring datasets to the needs of particular studies, streamlining data exploration and analysis. 

\begin{figure}[ht]
    \centering
    \includegraphics[width=0.8\textwidth]{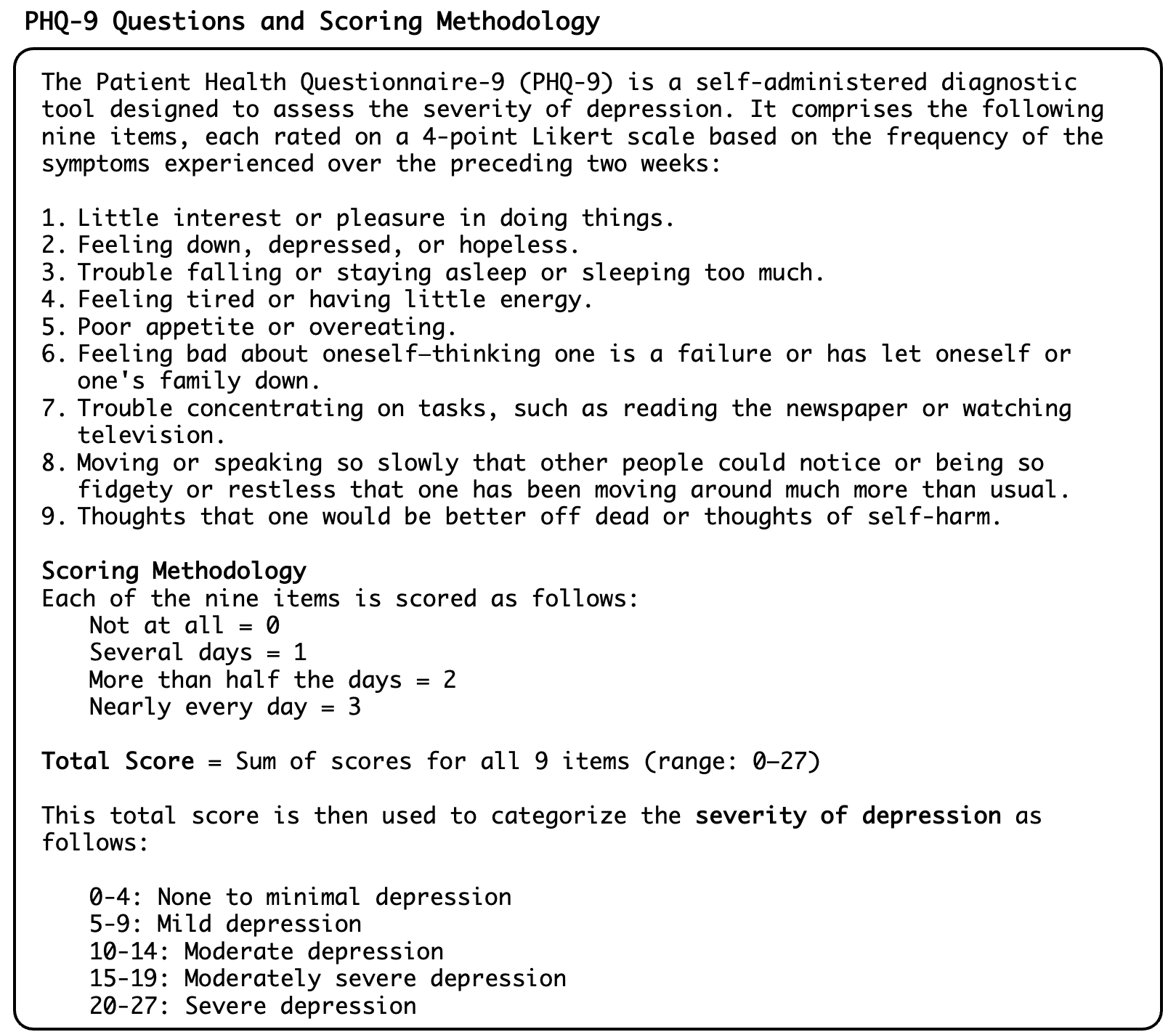}
    \caption{PHQ-9 Questionnaire and Risk Score Calculation: This figure illustrates the nine-item Patient Health Questionnaire (PHQ-9) with corresponding scoring criteria. Each item is rated from 0 (Not at all) to 3 (Nearly every day), and the total score is used to assess depression severity.}
    \label{fig:phq9_questions_score}
\end{figure}

The functions within this module are tailored to the type of resource being processed (Table \ref{tab:processing_functions}).
For example: clinicians and researchers often employ clinical questionnaires to derive risk scores relevant to their purpose. The Data Processing module includes built-in functions that automatically generate risk scores based on the particular survey, with calculation formulas for certain questionnaire types provided as part of the package.
Additionally, the module allows developers to implement custom risk score calculations if alternative methodologies or new questionnaires are required. For instance, a commonly used clinical questionnaire is the Patient Health Questionnaire-9 (PHQ-9) \cite{kroenke2001phq} (illustrated in Figure~\ref{fig:phq9_questions_score}). The Spezi Data Pipeline supports automated extraction and scoring of PHQ-9 responses and similarly structured questionnaire data.

\begin{table}[h!]
    \centering
    \begin{tabular}{|c|c|p{6cm}|}
        \hline
        \textbf{Function} & \textbf{Resource Type} & \textbf{Description} \\ \hline
        Outliers filtering & Observation & Filters out non-representative and abnormal data values. \\ \hline
        User(s) selection & Observation & Selects and returns data for specific users based on the user ID. \\ \hline
        Date(s) selection & Observation & Filters and returns data within a specified date range. \\ \hline
        Daily data aggregation & Observation & Aggregates daily data, summing values within a day. \\ \hline
        Daily averaging & Observation & Calculates the daily average of the data. \\ \hline
        Activity index calculation & Observation & Calculate the 7-day moving average of step counts. \\ \hline
        Risk score calculation & QuestionnaireResponse & Calculates the risk score based on the questionnaire responses. \\ \hline
    \end{tabular}
    \caption{Functions in data processing module.}
    \label{tab:processing_functions}
\end{table}

\textbf{Data Exploration.} The Data Exploration module provides tools for visualizing various types of data, supporting both individual and combined datasets.
This module offers different visualization options depending on the nature of the data (\eg~ bar plots, scatter plots, line charts), including specialized tools for plotting observations and \gls{ECG}s. Designed for healthcare professionals, researchers, and data analysts, the module extends foundational visualization capabilities, introducing advanced functionalities such as individual \gls{ECG} lead visualization, date and user filtering, and customizable plotting aesthetics.
The Data Exploration module additionally includes functions for plotting and analyzing variations in risks scores derived from stored questionnaire responses, aiding in the identification of trends in patient-reported outcomes.

\textbf{Data Export.} The Data Export module enables users to export data and visuals in various formats.
Tables can be exported to \textit{.CSV}/\textit{.XLSX} formats, and plots and figures to \textit{.PNG}/\textit{.TIFF}/\textit{.JPEG} formats.
This module ensures that processed and visualized data can be easily shared and applied across diverse use cases.
Notably, tabular data is optimized for machine learning applications, ensuring data integrity, standardization, and basic processing.
This capability supports the integration of findings into broader data analysis workflows, enhancing the reproducibility, scalability, and impact of digital health studies.

\subsection{System Evaluation}
To evaluate the Spezi Data Pipeline's utility in a real-world clinical research setting, we deployed it within the ongoing Pediatric Apple Watch Study (PAWS) at Stanford University. This implementation involved managing diverse digital health observations collected from mobile applications and wearable devices, including measurements such as heart rate, daily step counts, estimated energy expenditure (calories burned), levels of physical activity or exertion, and maximum oxygen uptake ($\mathrm{VO}_{2}\ \max$), an indicator of cardiovascular fitness. All data were structured according to the FHIR standard, preserving their hierarchical format with proper use of standardised coding systems (\eg, LOINC), timestamps, and patient identifiers. These FHIR resources were stored in Firebase Firestore, enabling the Pipeline's Data Access module to retrieve and process them efficiently in a cloud-native architecture.

\subsubsection{Pediatric Apple Watch Study: Real-World Validation}

Heart rhythm abnormalities are a leading cause of death in children with congenital heart disease.
Early detection and characterization of arrhythmias in pediatric patients can be critical for developing treatment strategies that improve survival rates and overall quality of life.
Children presenting with arrhythmia symptoms typically require ambulatory monitoring to determine the etiology of their symptoms. However, existing medical grade monitors often have limited wear times, bulky designs, and irritating adhesives, rendering them impractical for long-term use. 

Convincing evidence supports the use of the Apple Watch for arrhythmia characterization in adults, but data on its efficacy in pediatric populations remain limited. In recognition of this gap, the Pediatric Apple Watch Study (PAWS) at Lucile Packard Children's Hospital at Stanford University--a prospective, single-center observation study--aims to determine whether the Apple Watch can be used as a cardiac event monitor for children by comparing Apple Watch \gls{ECG} recordings and traditional ambulatory cardiac monitors in pediatric patients undergoing clinical arrhythmia evaluation.

The PAWS mobile application was built using the modular Spezi digital health ecosystem. Leveraging Spezi modules allowed seamless integration with the Apple Watch, enabling efficient data collection and secure transfer to Firebase. Pediatric patients aged 6 to 18 years who were undergoing clinically indicated arrhythmia monitoring with a Zio patch, 24-48 hour Holter monitor, or event monitor recording device were enrolled in the study. Participants were required to possess the developmental ability, as determined by both the patient and their parents, to safely wear the Apple Watch and utilize the patient-activated trigger functions on both the clinical monitor and the Apple Watch.

Patients younger than 6 years of age, those unable to effectively utilize the triggered features of either device, or those unable to use the Apple Watch were excluded from participation.
All eligible patients were offered participation during routine clinical visits.
Study participants were given an Apple Watch to be worn for a duration of 6 months and instructed to record an Apple Watch \gls{ECG} tracing each time they felt symptoms concerning for an arrhythmia.
The study protocol was approved by the Stanford University Institutional Review Board (Protocol \#67366), and all procedures adhered to institutional and federal ethical guidelines.

In order to effectively compare the Apple Watch against current gold-standard monitors, a data visualization dashboard was developed with Spezi Data Pipeline to enable physicians to analyze incoming Apple Watch data alongside recordings from standard monitors.
The resulting dashboard tracks whether \gls{ECG}s have been reviewed by clinicians and allows annotations on tracing quality, \gls{ECG} diagnoses, and reviewer comments, ensuring a comprehensive evaluation of the Apple Watch's performance in detecting and characterizing arrhythmias in children.
The following steps detail the the process of leveraging the Pipeline to establish an efficient method for accessing cloud-stored study data and preparing them for later analysis and review.  

The first step involved accessing the \gls{FHIR} observations stored in Firebase Firestore, which included step counts, heart rate, and \gls{ECG} recordings.
The Spezi Data Pipeline was configured to connect to the Firebase Firestore server, using secure authentication protocols to ensure data privacy and security. The data of interest-including \gls{ECG} recordings stored as FHIR observations-were specified through the Pipeline's interface.
Upon retrieval, the \gls{ECG} data were processed and transformed into a structured flat-table format, including fields such as patient ID, resource ID, timestamp, heart rate, \gls{ECG} waveform data, and Apple Watch-provided classifications.
These classifications helped to facilitate the identification and prioritization of abnormal recordings for review.
The resulting table was exported locally for the study's records and then fed into the software's exploration tools.

Two workflows were developed in order to load, pre-process, and evaluate the stored \gls{ECG} data.
These workflows incorporated the open-source Spezi Data Pipeline alongside an additional data review component, enabling clinicians to assign diagnoses and annotate recordings with relevant notes.
A custom-built interactive tool was specifically designed for reviewing \gls{ECG} data.
This tool integrates the visualization capabilities of the Spezi Data Pipeline into an interactive dashboard in the form of a Python notebook.
Clinicians access the system by connecting to the database, downloading the required data, and utilizing the tool to load and evaluate \gls{ECG} recordings.
Clinicians must enter their initials to initiate the review process, after which the dashboard displays the number of recordings available for review.
For each \gls{ECG} trace, clinicians can assign a diagnosis from predefined categories, including "Normal Sinus Rhythm," "Sinus Tachycardia," "Supraventricular Tachycardia (SVT)," "Ectopic Atrial Tachycardia (EAT)," "Atrial Fibrillation (AF)," "Ventricular Tachycardia (VT)," "Heart Block," or "Other." 
Clinicians can additionally assess trace quality using a five-tier scale ("Uninterpretable," "Poor Quality," "Adequate," "Good," or "Excellent"), and have the option to add patient- or trace-specific notes to provide further context or clarification.
All annotations and diagnoses are saved and stored back to the cloud. A schematic representation of the interactive dashboard is shown in Figure \ref{fig:paws_workflow} (top panel). 

Additionally, an alternative workflow is available, enabling clinicians to navigate through the \gls{ECG} recording and allowing the selection and visualization of specific recordings using various filters, including age group, \gls{ECG} classification (as provided by the AppleWatch), user ID, and date.
The notebook-based interactive dashboard is illustrated in Figure \ref{fig:paws_workflow} (bottom panel). 
\section{Results}

\begin{figure}[h!]
  \centering
  \includegraphics[width=\linewidth]{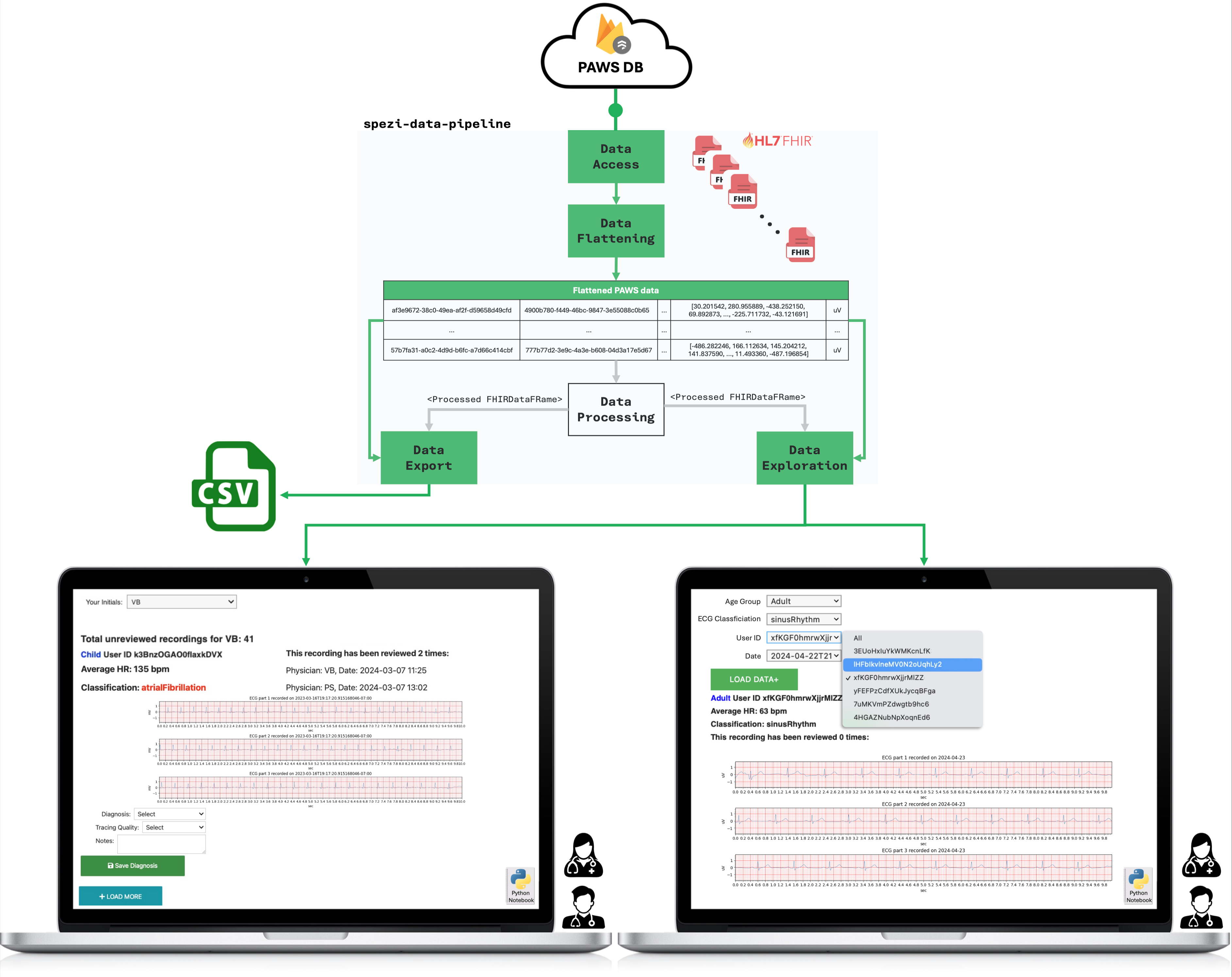}
  \caption{Overview of the adoption of the Spezi Data Pipeline for PAWS: The PAWS data stored on Firebase Firestore is accessed and fetched by the data access module. The Spezi Data Pipeline visualization tools are integrated in a Python notebook allowing clinicians to review, annotate, and compare ECG recordings. The figure contains test data for demonstration purposes.}
  \label{fig:paws_workflow}
\end{figure}

The Spezi Data Pipeline was implemented in the Pediatric Apple Watch Study (PAWS) to evaluate its ability to streamline access, processing, and visualization of digital health data stored in Firebase Firestore (Figure \ref{fig:paws_workflow}). The Pipeline successfully enabled secure retrieval and transformation of ECG recordings and related health metrics into a structured, compressed tabular format, which could be exported locally for analysis.

Apple Watch ECG recordings from the pediatric participants enrolled in PAWS were analyzed. At the time of analysis, 100+ subjects (age range: 6.1 - 18 years) were enrolled in PAWS, and a total of 4,000+ electrocardiogram (ECG) recordings had been collected.

\begin{figure}[htbp]
    \centering
    \includegraphics[width=\linewidth]{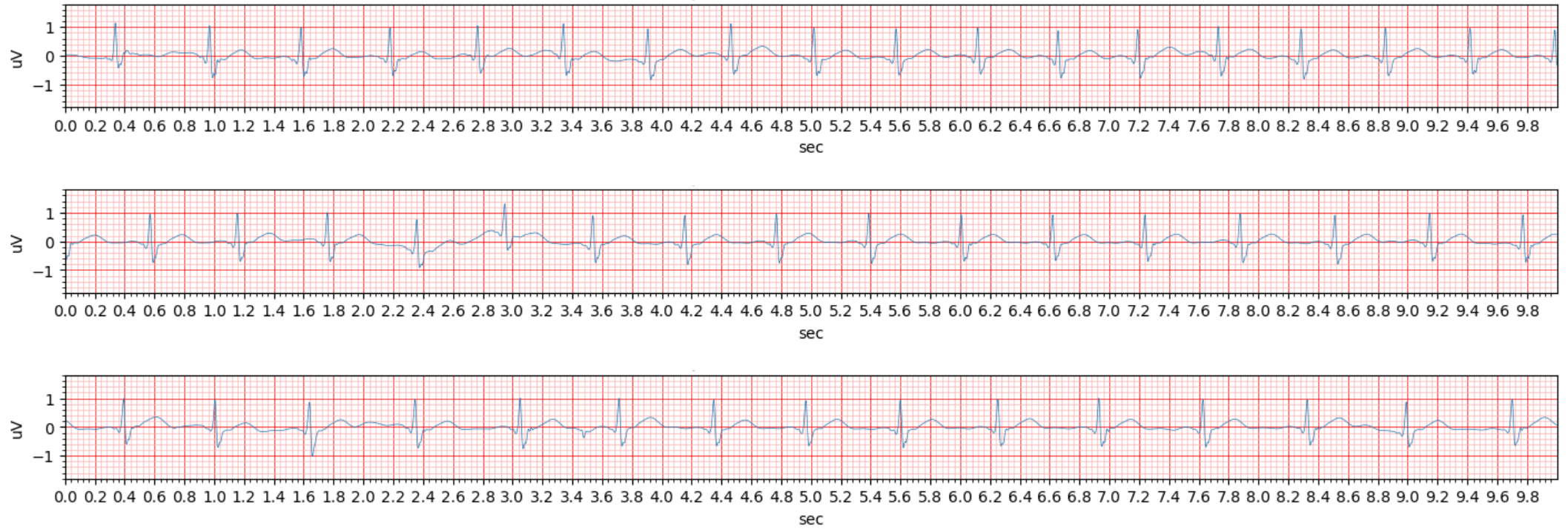}
    \caption{\textbf{Representative normal ECG from a pediatric participant.} 
    The tracing demonstrates normal sinus rhythm with age-appropriate heart rate and intervals similar to a typical Lead I ECG tracing. The P waves are upright, the PR interval is within normal limits, and QRS complexes are narrow and upright, consistent with normal conduction in a healthy pediatric heart.}
    \label{fig:pediatric-normal-ecg}
\end{figure}

Representative cases demonstrate the utility of the Spezi Data Pipeline in visualizing and exploring ECG recordings collected from pediatric participants. Figure~\ref{fig:pediatric-normal-ecg} presents a normal sinus rhythm tracing with an average heart rate of 66 bpm. Through the Pipeline's standardized data access and visualization modules, the tracing can be easily inspected to reveal upright P waves, narrow QRS complexes, and normal PR intervals, consistent with typical pediatric conduction patterns.

\begin{figure}[htbp]
    \centering
    \includegraphics[width=\linewidth]{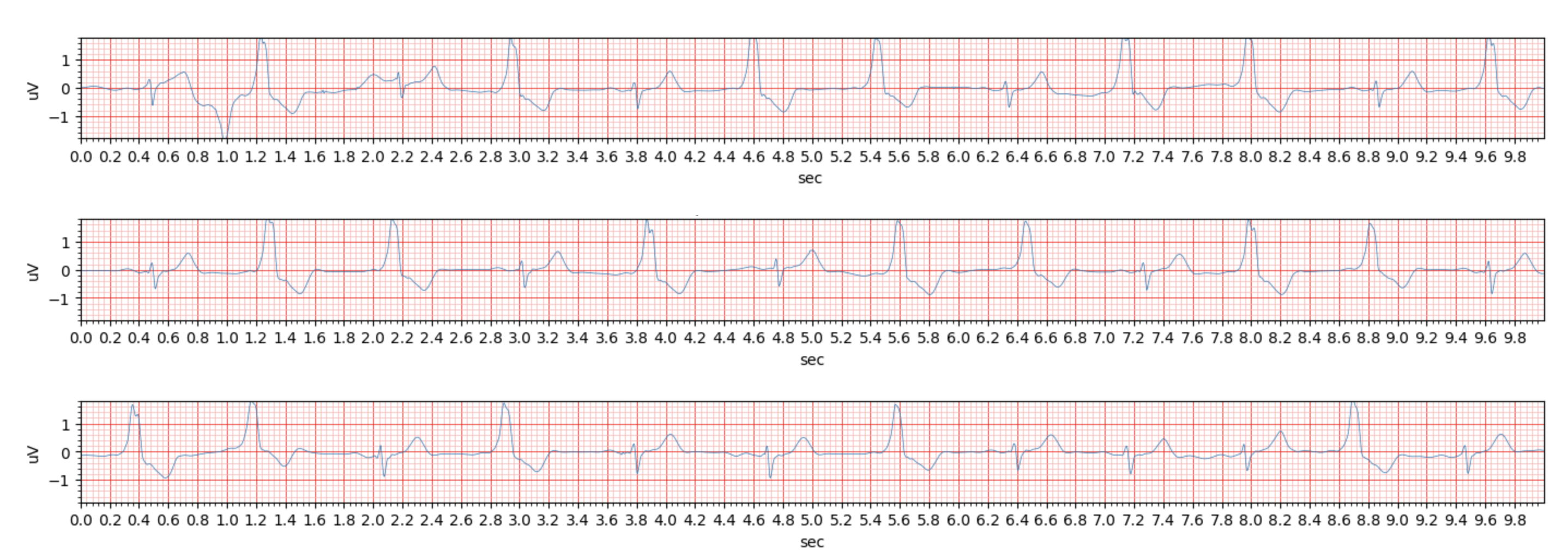}
    \caption{\textbf{Representative ECG tracing from a pediatric participant with Wolff-Parkinson-White (WPW) syndrome.} 
    The participant's baseline heart rate was 66 bpm.  The tracing shows sinus rhythm with intermittent wide QRS complexes, consistent with intermittent preexcitation. These findings are characteristic of WPW syndrome and illustrate the presence of an accessory pathway.}
    \label{fig:pediatric-wpw-ecg}
\end{figure}

Figure~\ref{fig:pediatric-wpw-ecg} shows a baseline ECG from another participant with Wolff-Parkinson-White (WPW) syndrome. The Pipeline enables clear visualization of intermittent pre-excitation, including widened QRS complexes and short PR intervals, patterns that are readily identifiable through harmonized rendering of the signal data.

\begin{figure}[htbp]
    \centering
    \includegraphics[width=\linewidth]{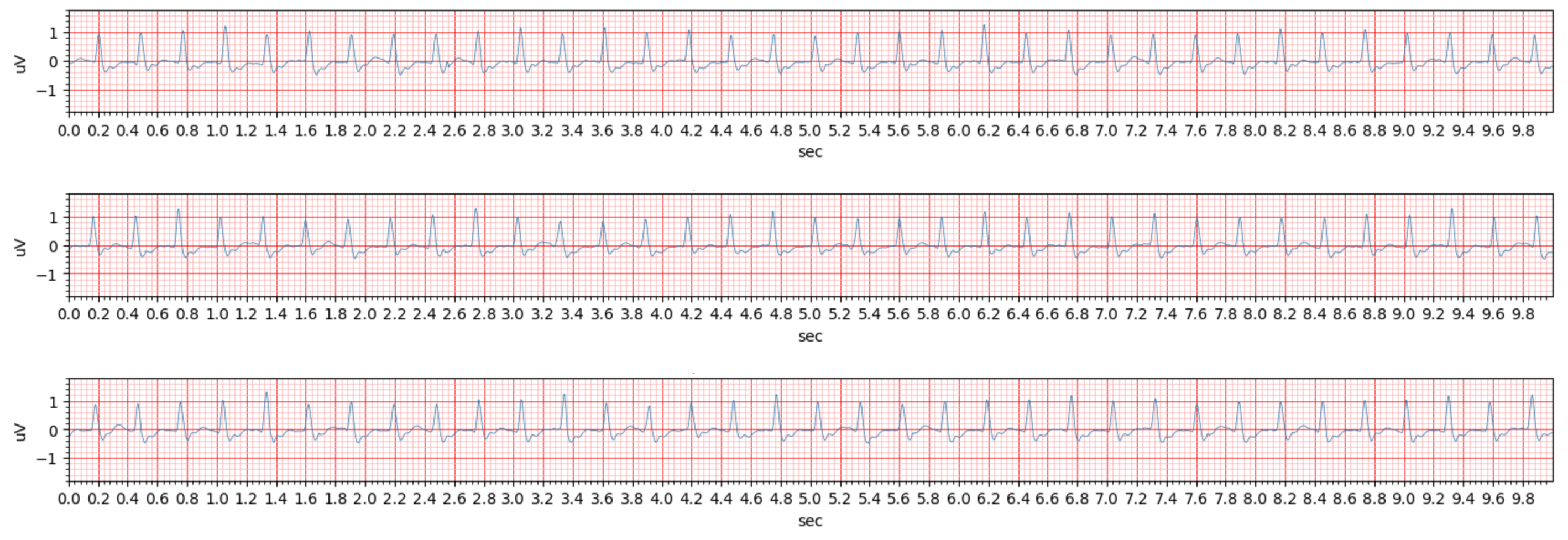} 
    \caption{\textbf{ECG tracing from a pediatric participant demonstrating supraventricular tachycardia (SVT).} 
    The single lead ECG shows a regular, narrow complex tachycardia consistent with SVT. The ECG tracing was captured during a symptomatic episode of palpitations and demonstrated SVT at a rate of 190 bpm.}
    \label{fig:pediatric-svt-ecg}
\end{figure}

In Figure~\ref{fig:pediatric-svt-ecg}, a third participant experienced an episode of supraventricular tachycardia (SVT) with a heart rate of approximately 190 bpm. The Spezi Data Pipeline enables detailed visualization of this event, capturing a narrow complex, regular tachycardia in a symptomatic child during palpitations and demonstrating the platform's ability to support arrhythmia identification in pediatric patients.

Collectively, these cases highlight how the Pipeline supports reproducible exploration of wearable ECG data, enabling researchers and clinicians to visualize normal conduction, baseline abnormalities, and transient arrhythmic events within a unified analytic environment. Rather than relying on device-specific outputs, the Spezi Data Pipeline provides a flexible framework for inspecting, comparing, and interpreting signals in a research context.

Visualization tools integrated into the interactive dashboard (integrated in a Python notebook) allowed clinicians to efficiently review, annotate, and compare ECG traces, with all diagnoses and quality assessments saved back to the cloud. 
Throughout the process, patient identifiers remained encoded to protect privacy, and access to identifiable information was restricted to valid clinical scenarios. 
Overall, integration of the Spezi Data Pipeline provided an efficient and robust solution for data processing, exploration, and visualization within the PAWS digital health study. 
The Pipeline enabled seamless extraction, structuring, and analysis of complex health data, significantly improving workflow efficiency and supporting clinician-driven review while maintaining high standards of data security and interoperability.

\begin{figure}[ht]
    \centering
    \includegraphics[width=0.85\textwidth]{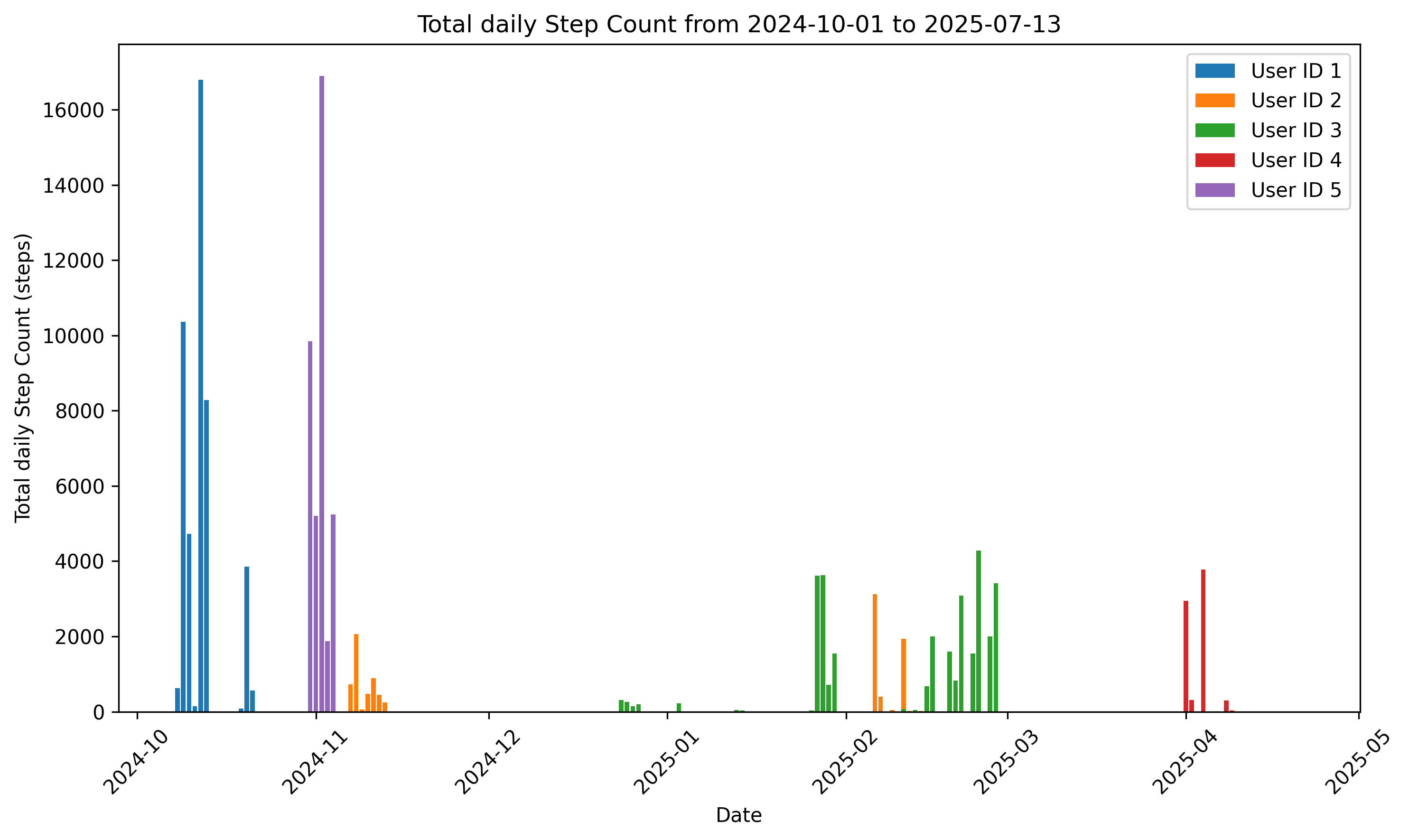}
    \caption{
        \textbf{Total daily step counts for five indicative users.}
        Each line represents the total number of steps recorded per day for each user over the observation period.
    }
    \label{fig:daily_steps_five_users}
\end{figure}

\begin{figure}[ht]
    \centering
    \includegraphics[width=0.85\textwidth]{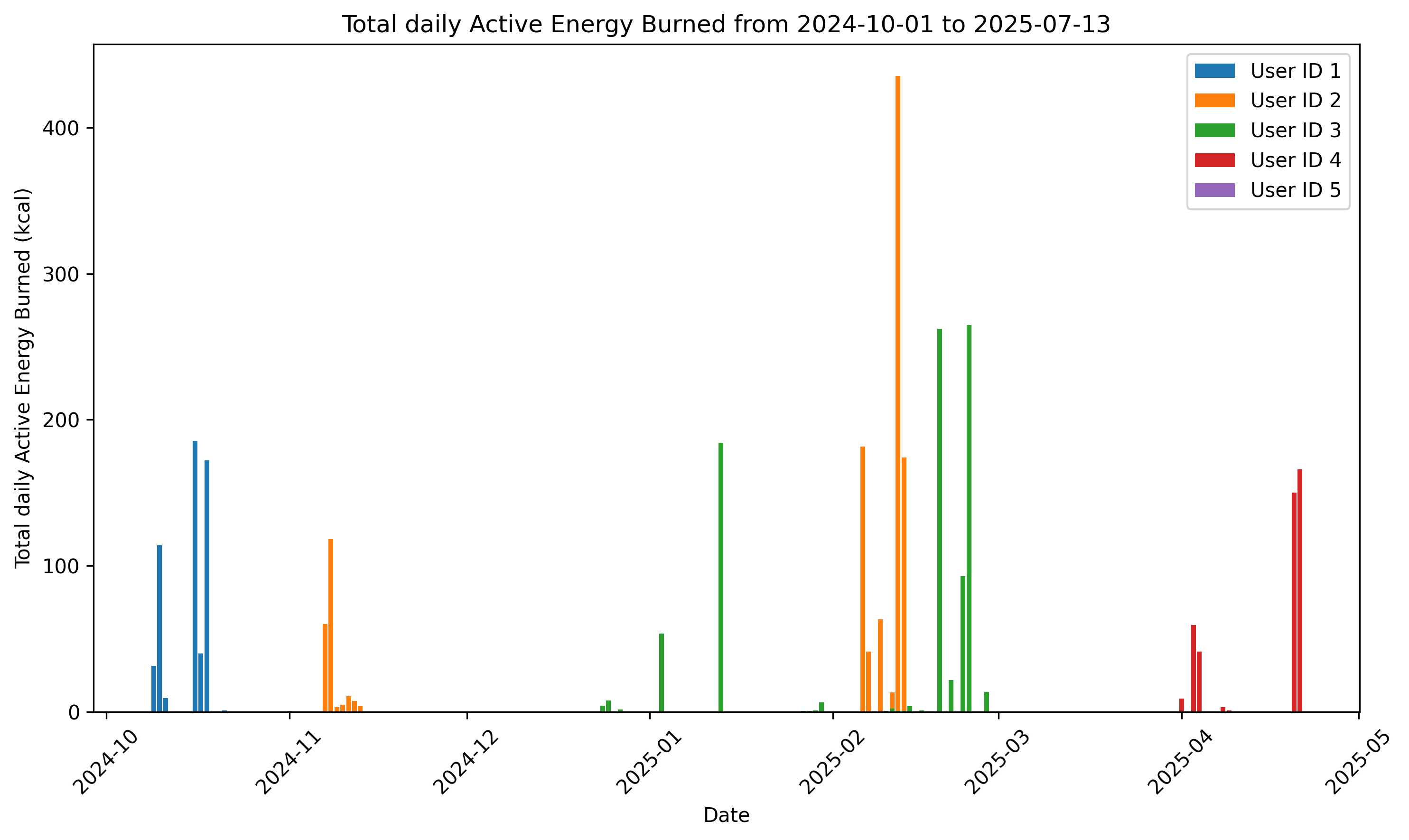}
    \caption{
        \textbf{Total daily calories burned for five indicative users.}
        The figure shows the total calories burned each day for the same set of users.
    }
    \label{fig:daily_calories_five_users}
\end{figure}

\begin{figure}[ht]
    \centering
    \includegraphics[width=0.85\textwidth]{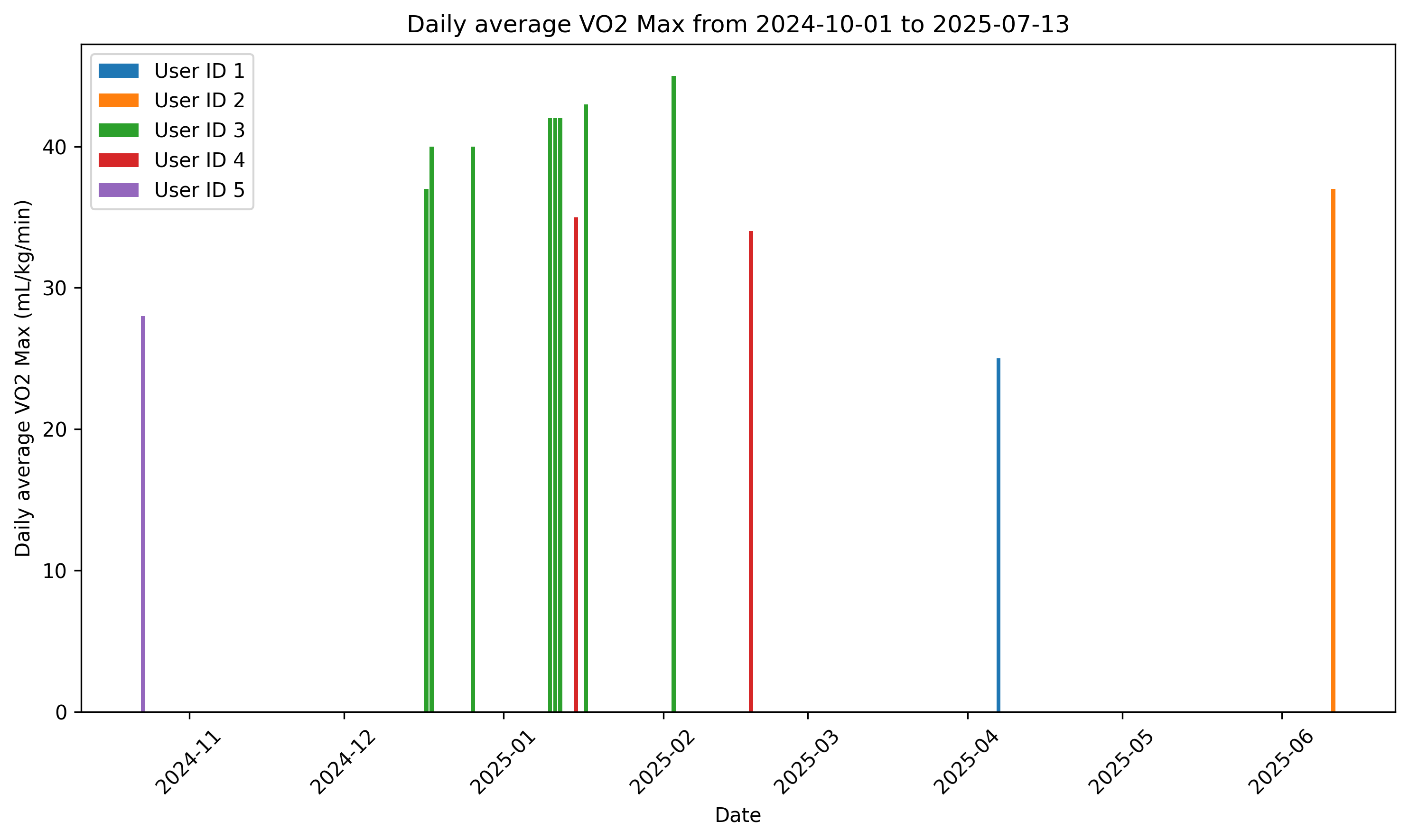}
    \caption{
        \textbf{Daily average VO\textsubscript{2}max for five indicative users.}
        The plot presents the daily mean VO\textsubscript{2}max values for each user.
    }
    \label{fig:daily_vo2max_five_users}
\end{figure}

\begin{figure}[ht]
    \centering
    \includegraphics[width=0.85\textwidth]{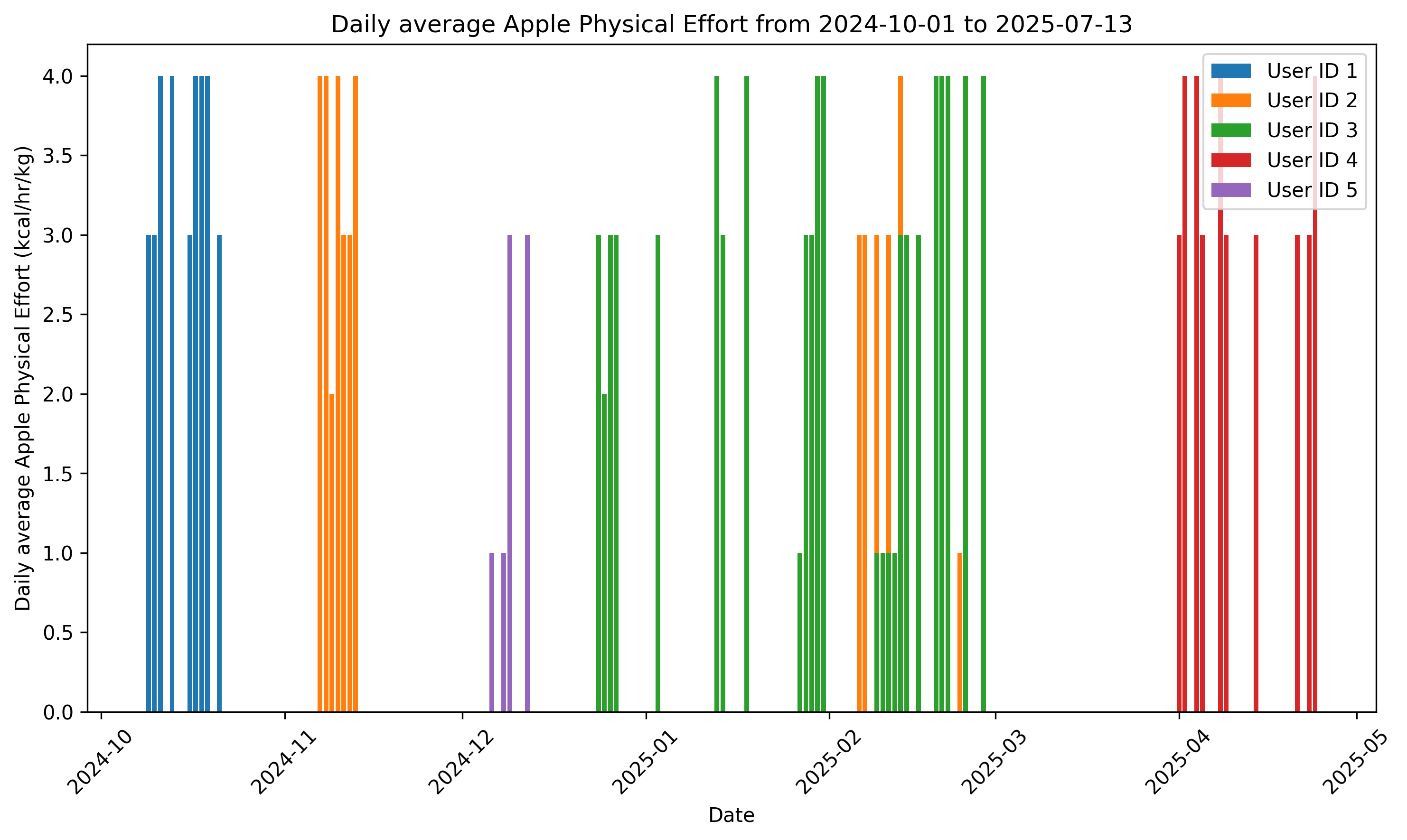}
    \caption{
        \textbf{Daily average physical effort for five indicative users.}
        This figure shows the daily average physical effort, as estimated by the wearable device, for each user.
    }
    \label{fig:daily_physical_effort_five_users}
\end{figure}

\begin{figure}[ht]
    \centering
    \includegraphics[width=0.85\textwidth]{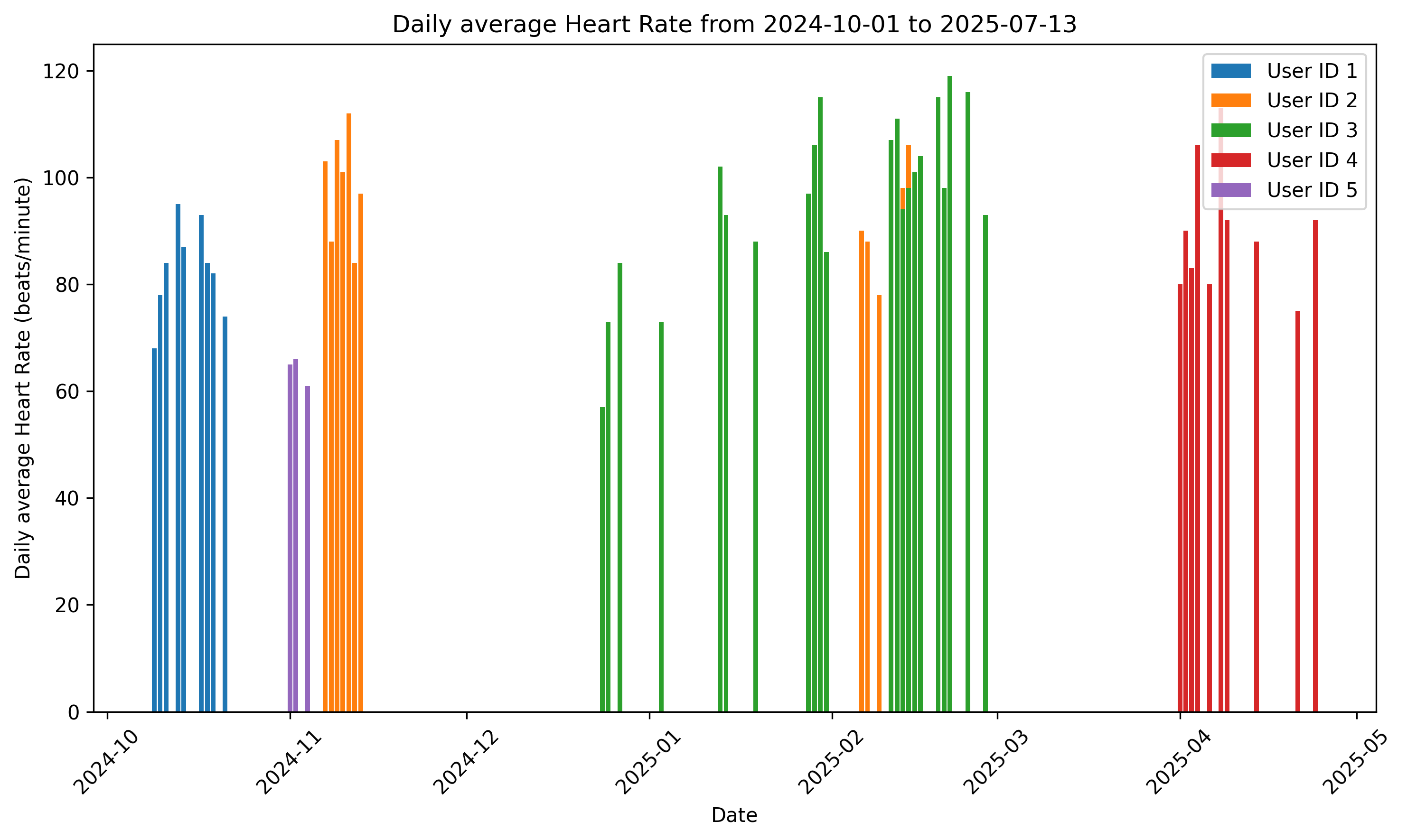}
    \caption{
        \textbf{Daily average heart rate for five indicative users.}
        The plot presents the mean heart rate recorded per day for each user. 
    }
    \label{fig:daily_average_heart_rate_five_users}
\end{figure}

\begin{figure}[ht]
    \centering
    \includegraphics[width=0.95\textwidth]{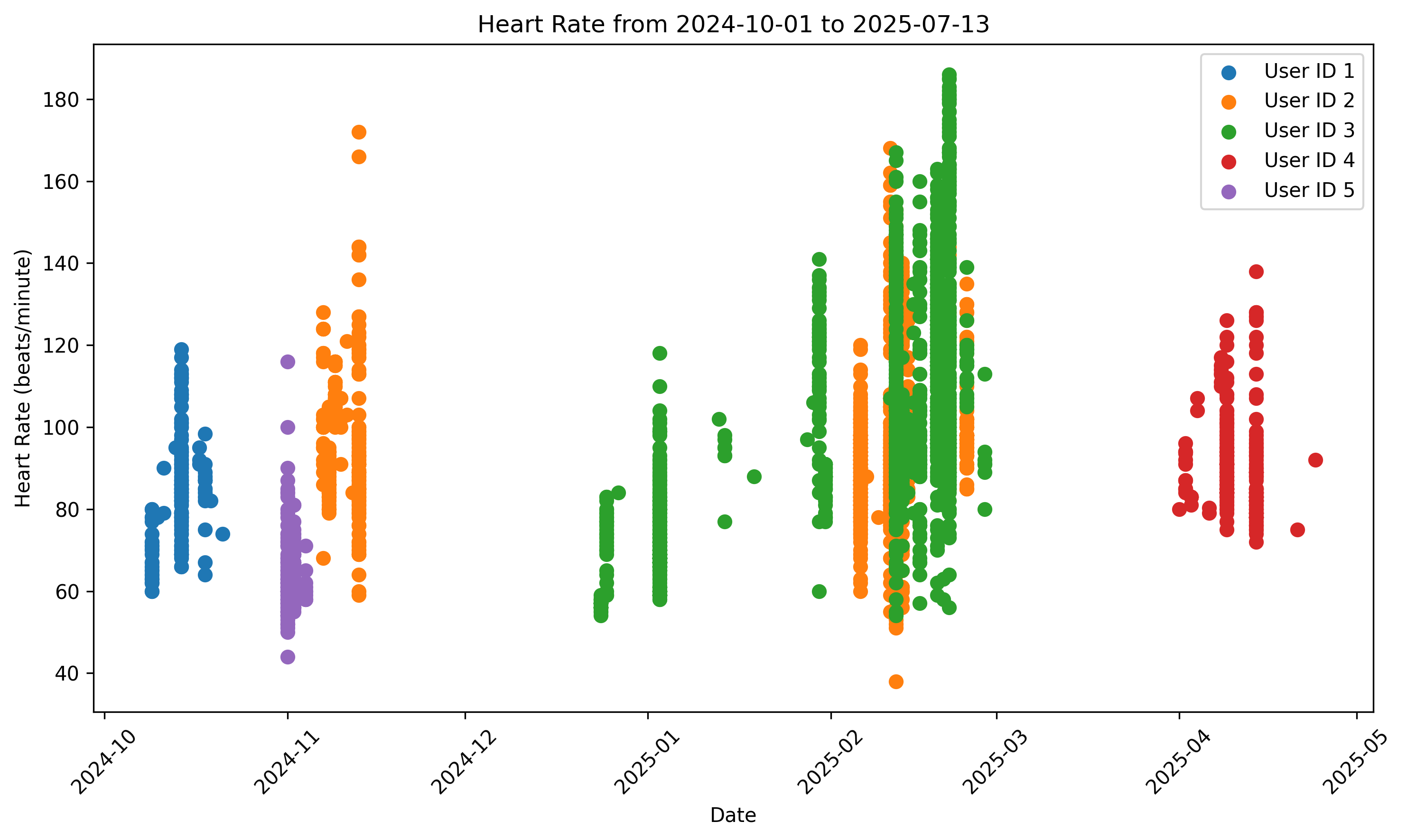}
    \caption{
        \textbf{Heart rate data distributions for five indicative users.}
        The plot presents the distribution of all heart rate measurements (dots) for each user. Multiple heart rate recordings per day are visualized to illustrate intra-day variability.
    }
    \label{fig:daily_heart_rate_distribution_five_users}
\end{figure}

Furthermore, the Spezi Data Pipeline data processing and data exploration modules enabled the seamless acquisition, harmonization, and visualization of diverse physiological measurements, including step counts, calories burned, VO$_2$max, physical effort, and heart rate, across multiple users. This streamlined approach facilitated rapid data filtering and aggregation, while automated identifier masking ensured robust patient privacy protection throughout the analytic workflow. The integrated functionalities of the Spezi Data Pipeline enabled automatic aggregation of either average or total daily values for each physiological measurement (Figures~\ref{fig:daily_steps_five_users}, \ref{fig:daily_calories_five_users}, \ref{fig:daily_vo2max_five_users}, \ref{fig:daily_physical_effort_five_users}, \ref{fig:daily_average_heart_rate_five_users}); when appropriate, the full distribution of multiple daily measurements for each user was visualized (Figure~\ref{fig:daily_heart_rate_distribution_five_users}). Five representative users were arbitrarily selected for visualization purposes. The tool provides flexibility to select specific user IDs and date ranges for targeted investigation. User identifiers were masked throughout to ensure privacy.

\begin{figure}[h]
    \centering
    \includegraphics[width=0.7\textwidth]{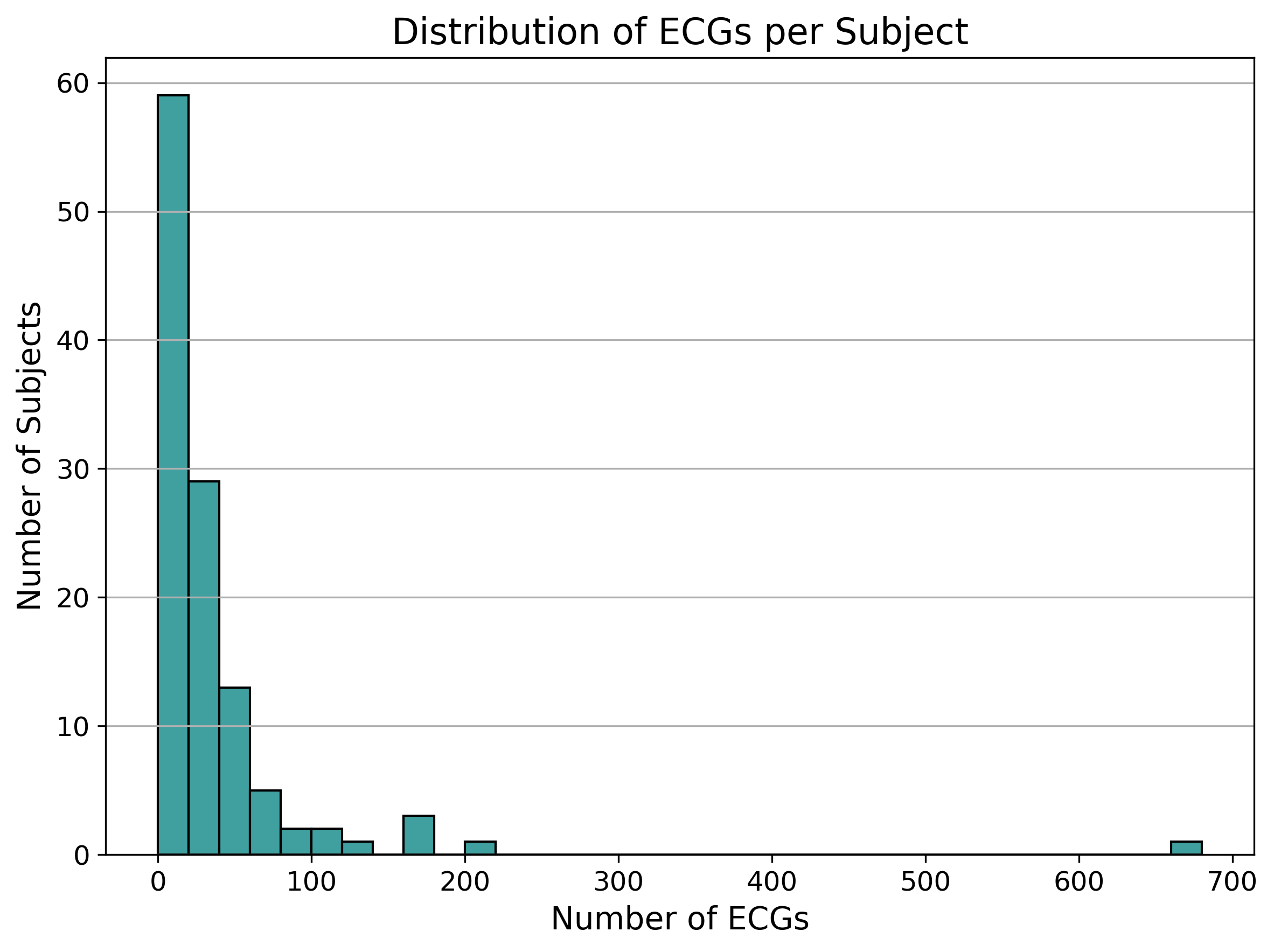}
    \caption{Number of ECG recordings per subject}
    \label{fig:ecg_counts_per_subject}
\end{figure}

Complementing the above, Figure~\ref{fig:ecg_counts_per_subject} demonstrates how the Spezi Data Pipeline can be used to quantify and visualize the number of ECG recordings contributed by each participant. The distribution highlights heterogeneity in monitoring frequency and engagement, with most participants providing fewer than 100 recordings, while a subset contributed substantially more.

\begin{figure}[h]
    \centering
    \includegraphics[width=0.7\textwidth]{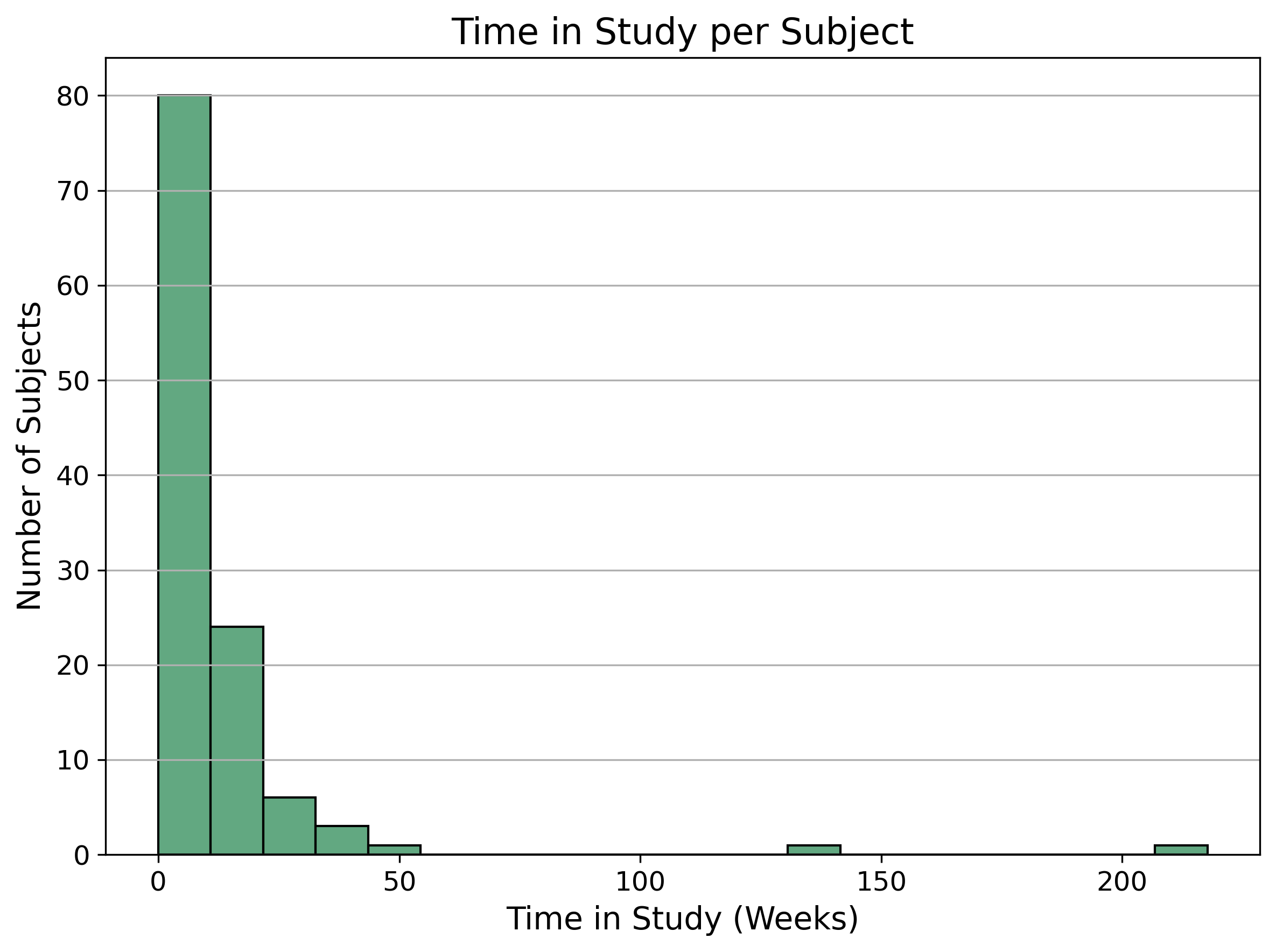}
    \caption{Time in study per subject (weeks)}
    \label{fig:time_in_study}
\end{figure}

To further contextualize study participation, Figure~\ref{fig:time_in_study} illustrates participant follow-up duration, measured in weeks, using the Pipeline's visualization framework. While the majority of subjects were active for 10 weeks or less, a substantial subset remained engaged for over 50 weeks. These views underscore how the Pipeline supports exploration of study dynamics, allowing researchers to assess variability in data contribution and retention across individuals in a longitudinal context.


\section{Discussion}

This study introduces the Spezi Data Pipeline, an open-source Python package developed to address critical ongoing challenges in interoperability, scalability, and data integrity in digital health data management.
By adhering to HL7 FHIR standards, the Pipeline ensures that diverse health data--from sensor-derived observations to clinical questionnaire responses--can be securely and consistently handled across research and clinical environments.
The modular architecture (comprising of data access, flattening, processing, exploration, and export modules) enables flexible integration into existing workflows and supports a broad range of analytical and visualization tasks.

The Spezi Data Pipeline enables systematic, modular testing across all core components, simulating diverse user interactions and health metrics to rigorously validate functionality and reliability. Its ability to transform complex, hierarchical FHIR resources into flat, analysis-ready tables is essential for efficient downstream data exploration and modeling, supporting rapid and reproducible analytics. Integrated visualization tools provide clear, actionable insights into activity, nutrition, vitals, and risk scores, demonstrating the platform's utility for both clinical research and operational decision-making.

The Pipeline's real-world utility was further validated through its deployment in the Pediatric Apple Watch Study (PAWS) at Stanford University.
Integration with Firebase Firestore enabled secure, FHIR-compliant retrieval and transformation of ECG and related health data, ensuring data integrity and interoperability throughout.
The interactive dashboard, developed on top of the Pipeline, facilitated efficient clinician review, annotation, and comparison of ECG traces, with all actions securely logged and synchronized to the cloud. 
Over 4,000 ECGs have been recorded by study participants and reviewed using this Pipeline.
This workflow enhanced the efficiency and reproducibility of our digital health study while maintaining strict data privacy and security standards through the use of encoded patient identifiers and controlled access to sensitive information.

Despite widespread adoption of FHIR in clinical systems, few research-focused tools offer seamless end-to-end pipelines for transforming FHIR-based health data into analysis-ready formats. The Spezi Data Pipeline fills this infrastructure gap by combining standards-based design with accessible, open-source tooling, lowering the barrier for health data science. This design not only supports efficient reuse of structured health data across studies, but also advances the reproducibility and transparency goals central to open science. By automating the preprocessing of FHIR resources--often a time-consuming and error-prone task--Spezi enables more consistent benchmarking and collaborative research on digital health datasets.

The Pipeline's modular architecture enables seamless integration with existing systems while maintaining strict compliance with interoperability standards, allowing research teams to focus on scientific questions rather than technical implementation details. 
By streamlining FHIR-compliant workflows and enabling structured expert review, the Spezi Data Pipeline lowers technical barriers, reduces resource demands, and facilitates the translation of complex health data into actionable insights. 

\subsection*{Limitations and Future Directions}
Although the Spezi Data Pipeline is designed as a generalizable, modular framework, the scope of the present study was limited to physiological metrics and questionnaire responses collected within a single institution. Evaluations across other supported resource types, diverse clinical settings, populations, and health systems will be critical to assessing the Pipeline's cross-contextual utility.

Future efforts include ecosystem expansion through additional modular components, enhanced support for diverse server connections, and alignment with complementary initiatives, such as JupyterHealth. As an open-source framework, the Spezi Data Pipeline is designed to be extensible and community-driven, encouraging researchers and developers to contribute new capabilities tailored to their domain-specific needs. These continued efforts will further strengthen the foundation for interoperable, reproducible, and scalable digital health research-ultimately enabling more efficient study designs and improved patient outcomes.

\section*{Data Availability}
The real-world data from the Pediatric Apple Watch Study (PAWS) is not publicly available due to patient privacy considerations and regulatory restrictions. Access to PAWS data is governed by Stanford University Institutional Review Board (IRB) protocols and may be granted upon request, subject to appropriate ethical approvals and data use agreements.

\section*{Code Availability}
The Spezi Data Pipeline codebase is available on \href{https://github.com/StanfordSpezi/SpeziDataPipeline}{GitHub}. It is also distributed as a Python package via PyPI: \url{https://pypi.org/project/spezi-data-pipeline/0.1.0/}. Documentation, usage examples, and contribution guidelines are provided in the GitHub repository. An experimental codebase demonstrating how to use the \texttt{spezi\_data\_pipeline} package for managing, analyzing, and visualizing healthcare data from Firebase Firestore is available at \url{https://github.com/StanfordSpezi/SpeziDataPipelineTemplate}. This template showcases practical examples and use cases to support integration into real-world research workflows.

\section*{Acknowledgments}
We gratefully acknowledge the support of the Stanford Mussallem Center for Biodesign, Lucile Packard Children's Hospital, and the Departments of Pediatrics (Division of Pediatric Cardiology) and Surgery at Stanford University. Their institutional support was essential to the design and execution of this study.

\bibliographystyle{naturemag} 
\bibliography{bibliography.bib}

\end{document}